\documentclass[10pt,twocolumn,prl,showpacs,floatfix]{revtex4}
\usepackage{dcolumn}                 
\usepackage{times}
\usepackage{nicefrac}
\usepackage{amsmath}
\usepackage{amsfonts}
\usepackage{amssymb}
\usepackage{amsthm}

\newcommand*{\cent}[1]{\multicolumn{1}{c}{$#1$}}
\newcolumntype{w}[1]{D{.}{.}{#1}}

\newcommand{\icm}{\mathrm{cm}^{-1}}

\begin{document}
\preprint{Version 4.0}

\title{Testing quantum electrodynamics in the lowest singlet states of beryllium atom}

\author{Mariusz Puchalski}
\affiliation{Faculty of Chemistry, Adam Mickiewicz University,
             Grunwaldzka 6, 60-780 Pozna{\'n}, Poland }

\author{Krzysztof Pachucki}
\affiliation{Faculty of Physics, University of Warsaw, Ho{\.z}a 69, 00-681 Warsaw, Poland}

\author{Jacek Komasa}
\affiliation{Faculty of Chemistry, Adam Mickiewicz University,
             Grunwaldzka 6, 60-780 Pozna{\'n}, Poland }

\date{\today}

\begin{abstract}
High-precision results are reported for the energy levels
of $2{^1S}$ and $2{^1P}$ states of the beryllium atom. Calculations are performed
using fully correlated Gaussian basis sets and taking into account the relativistic,
quantum electrodynamics (QED), and finite nuclear mass effects. Theoretical predictions
for the ionization potential of the beryllium ground state $75\,192.699(7)~\icm$ and the
$2{^1P} \rightarrow 2{^1S}$ transition energy  $42\,565.441(11)~\icm$
are compared to the known but less accurate experimental values. The accuracy of the four-electron 
computations approaches that achieved for the three-electron
atoms, which enables determination of the nuclear charge radii and precision tests of QED.
\end{abstract}

\maketitle

Spectroscopic standards for the energy transitions of the beryllium atom
have been established many years ago (1962) in the experiments by Johansson.
In most cases an accuracy of $0.01-0.02$ \AA \,\cite{johansson1962} has been
reached. The only more precise beryllium energy level determination comes from
the experiment by Bozman {\it et al.} \cite{bozman1953} in which the transition energy 
of $21\,978.925~\icm$ between
the ground and the $2\,^3 P_1$ state accurate to about $0.01~\icm$ ($0.002$ \AA \,)
has been measured. Seaton, through a fit to a collection of excited states data,
has determined the ionization potential (IP) of the ground state
to be $75\,192.56(10)~\icm$ \cite{seaton1976}. Later, this quantity has been
improved by Beigang {\it et al.}, who obtained $75\,192.64(6)~\icm$ \cite{beigang}.
Contemporary high precision calculations \cite{beqed,stankeS} are in good agreement
with the rather old experimental IP values. Nevertheless, the precision of the data 
available for beryllium is far from
being satisfactory compared to the exquisite accuracy of the modern atomic spectroscopy.
The level of the absolute precision achieved in modern measurements for three-electron systems
\cite{sansonetti2011,bushaw2007} is as many as four orders of magnitude higher than
that obtained in the case of beryllium. As it has been shown for two- and three-electron
atoms, the availability of such accurate data,
in connection with good understanding of the underlying atomic theory, opens up access to such
interesting applications like the determination of the nuclear charge radius or precision tests 
of the quantum electrodynamics (QED).

Remarkable advances in theoretical methods make it possible to approach
the spectroscopic accuracy for the energies and transition frequencies
of few-electron atoms. This challenge requires precise treatment of the electron
correlations as well as inclusion of relativistic and quantum electrodynamic effects.
The concise approach, which accounts for all the effects beyond the nonrelativistic
approximation, is based on the expansion of the energy levels in the fine structure
constant $\alpha$ (see Eq.~(\ref{eq:expansion}) below).
This method has been successfully applied in recent years to light atomic
and molecular systems \cite{yan_rel,lit_rel,komasa2011,przybytekHe2}.
The frontiers in this field of research have been established
by the calculation of higher order $(m\,\alpha^6)$ corrections to helium energy levels \cite{h6hel}
and of $m\,\alpha^7$ corrections to helium fine structure \cite{PY10}.

Up to now, the precision of theoretical predictions for the beryllium states 
with the non-vanishing angular momentum has been severely limited by the accuracy
of the lowest-order relativistic $(m\,\alpha^4)$ and QED $(m\,\alpha^5)$
corrections. The most accurate calculations including the relativistic
corrections was performed 20 years ago by Chung and Zhu \cite{chung1993}
at the full-core plus correlation level of theory,
whereas the QED effects have merely been approximated from hydrogenic formulas
\cite{drakeblH,drakelnk0}. This approach has turned out to be unsatisfactory
for it has led to a significant discrepancy between theoretical predictions
and experimental excitation energies. For instance, the theoretical result \cite{chung1993}
is by $3.45~\icm$ higher than the experimental value $2\,349.329(10)$ \AA\
of the $2{^1P} \rightarrow 2{^1S}$ transition energy
measured with $0.18~\icm$ uncertainty \cite{johansson1962}.
The fact that the theoretical excitation energy is higher than the experimental value
may indicate that correlation effects have not been incorporated satisfactorily.
Such a disagreement can only be resolved in an unequivocally more accurate
computation of nonrelativistic energies as well as the relativistic and QED effects using
the explicitly correlated wave functions. Recent nonrelativistic calculations
of low-lying $P$- and $D$-states with the relative precision of an order of $10^{-10}$-$10^{-11}$
\cite{bubinP,sharkeyD} represent a step in this direction.

In this paper, we present the first complete and highly accurate treatment
of the leading relativistic $(m\,\alpha^4)$ and QED $(m\,\alpha^5)$ effects
for a four electron  atomic $P$-state. We also improved
results for the $2^1S$ state, which permitted us to push the accuracy of the theoretical
predictions of the $2{^1P} \rightarrow 2{^1S}$ transition energy
beyond the experimental uncertainty. Additionally, in combination with the previously reported very accurate data
on beryllium cation \cite{lit_rel}, we obtained an improved ionization potential
with the accuracy an order of magnitude higher than that of the available experimental values.

In our approach, we expand the total energy not only in the fine structure constant $\alpha\approx 1/137$
but also in the ratio of the reduced electron mass to the nuclear mass $\eta=-\mu/m_N = -m/(m+m_N)\approx 1/16424$.
This way we reduce the isotope dependence to the prefactors only. In terms
of these two parameters, the energy levels can be represented as the following expansion
\begin{eqnarray}
E =&& m\,\alpha^2\,\bigl[{\cal E}^{(2,0)}+\eta\,{\cal E}^{(2,1)}\bigr] +\,m\,\alpha^4\,{\cal E}^{(4,0)}
\nonumber \\ && +m\,\alpha^5\,{\cal E}^{(5,0)} +\,m\,\alpha^6\,{\cal E}^{(6,0)} + \ldots\,.
\label{eq:expansion}
\end{eqnarray}
Each dimensionless coefficient ${\cal E}^{(m,n)}$ is calculated separately as an expectation value
of the corresponding operator with the nonrelativistic wave function.
The leading contribution ${\cal E}^{(2,0)}\equiv {\cal E}_0$ is an eigenvalue of the Schr{\"o}dinger
equation with the clamped nucleus Hamiltonian
\begin{eqnarray}
{\cal H}_0\Psi &=& {\cal E}_0 \Psi, \quad
{\cal H}_0 = \sum_a\frac{p_a^2}{2} -\sum_a\frac{Z}{r_a} +\sum_{a>b} \frac{1}{r_{ab}}.
\end{eqnarray}
The key to obtaining high-precision results is the use of a very accurate trial wave function $\Psi$,
which contains all the inter particle distances explicitly incorporated. We express $\Psi$ as a linear combination of $N$ four-electron basis functions $\psi_i$
\begin{eqnarray}
\Psi &=& \sum_i^N\,c_i\,\psi_i, \qquad \psi_i = {\cal A}[ \phi_i(\vec r_1,\vec r_2,\vec r_3,\vec r_4)\,\chi]\,,
\end{eqnarray}
where ${\cal A}$ is the antisymmetry projector, $\chi=\frac{1}{2}\left(\uparrow_{1}\downarrow_{2}-\downarrow_{1}
\uparrow_{2}\right)\left(\uparrow_{3}\downarrow_{4}-\downarrow_{3}
\uparrow_{4}\right)$ is the singlet spin function constructed using electron spinors.
The spatial function $\phi$ is the explicitly correlated Gaussian (ECG) function
for $S$- and $P$-state, respectively
\begin{eqnarray}
\phi_S &=&  \exp \big[-\sum_a w_a \,r^2_a -\sum_{a<b} u_{ab} \,r^2_{ab} \big],
\label{phiS} \\
\vec\phi_P &=& \vec r_1\,\exp \big[-\sum_a w_a \,r^2_a-\sum_{a<b} u_{ab} \,r^2_{ab} \big].
\label{phiP}
\end{eqnarray}
The main advantage of these Gaussian functions is the availability
of analytical forms of the integrals required for matrix elements
of the Hamiltonian ${\cal H}_0$
\begin{eqnarray}\label{eq:fintegral}
f(n_1,\ldots,n_{10}) &=& \int \ldots \int\frac{d^3 r_1}{\pi}\ldots \frac{d^3 r_4}{\pi} \,
                    r_{1}^{n_1}\ldots\,r_{4}^{n_4}  \\
  && \hspace{-2cm} \times \,r_{12}^{n_5}\ldots r_{34}^{n_{10}}\,
     \exp \big[-\sum_a \alpha_a \,r^2_a -\sum_{a<b} \beta_{ab} \,r^2_{ab} \big]\,. \nonumber
 \end{eqnarray}
Among all the integrals represented by the above formula we can distinguish
two subsets used in our calculations. The first subset contains the "regular"
integrals with the non-negative even integers $n_i$ such that $\sum_i n_i \leq \Omega_1$,
where the shell parameter $\Omega_1=0,2,4,\dots$.
The second subset permits a single odd index $n_i\geq-1$ for which
$\sum_i n_i \leq \Omega_{1/r}$ ($\Omega_{1/r}=-1,1,3,\dots$) and is related to
the components of the Coulomb potential.
To systematize the use of the ECG basis sets we re-derived the recurrence scheme
for the generation of both the classes of integrals from the master expression
\cite{gaussintegrals, harris}.
An advantage of such approach is the possibility of a gradual
extension of calculation to the states with higher angular momenta.
The sets of integrals employed in a specific case can be characterized
using the $\Omega$ shell parameters. For instance, the matrix elements
of the nonrelativistic Hamiltonian require integrals
with $\Omega_1=2$, $\Omega_{1/r}=-1$ for $S$-states (Eq.~(\ref{phiS})\,),
and $\Omega_1=4$, $\Omega_{1/r}=1$ for $P$-states (Eq.~(\ref{phiP})\,).
If, additionally, gradients with respect to the nonlinear parameters are to be used,
both shell parameters have to be increased by two.

To control the uncertainty of our results we performed the calculations with several basis sets
successively increasing their size by a factor of two. From the analysis of convergence 
we obtained the extrapolated
nonrelativistic energies and mean values of the operators presented in Table \ref{TBL1}.
The largest wave functions optimized variationally were composed of 4096 and 6144 terms
for the $S$- and $P$-state, respectively, leading to
the nonrelativistic energies ${\cal E}^{(2,0)}(2^1S) = -14.667\,356\,494\,9$~a.u. and
${\cal E}^{(2,0)}(2^1P) =  -14.473\,451\,33\,4$~a.u. These upper bounds improve slightly
those obtained by Adamowicz {\it et al.} \cite{stankeS,bubinP}.

The other coefficients of expansion \eqref{eq:expansion} are calculated
as mean values with the nonrelativistic wave function $\Psi$.
The nonrelativistic finite mass correction is given by
${\cal E}^{(2,1)} = {\cal E}^{(2,0)} - \sum_{a<b}\,\langle \vec p_a \cdot \vec p_b \rangle$.
In order to calculate the leading relativistic corrections ${\cal E}^{(4,0)} = \langle {\cal H}^{(4,0)} \rangle$
we consider the Breit-Pauli Hamiltonian \cite{bethe}, which for the states 
with vanishing spin can be effectively replaced by the form
\begin{eqnarray}
\label{eq:HBP}
{\cal H}^{(4,0)} &=& \sum_a \biggl[ -\frac{\vec p^{\,4}_a}{8}  + \frac{ \pi\,Z\,\alpha}{2}\,\delta^3(r_a)\biggr ]
 \\
&&  +\sum_{a<b} \biggl [\pi\, \delta^3(r_{ab})
-\frac{1}{2}\, p_a^i\,\biggl(\frac{\delta^{ij}}{r_{ab}}+\frac{r^i_{ab}\,r^j_{ab}}{r^3_{ab}}
\biggr)\, p_b^j \biggr].\nonumber
\end{eqnarray}
Since the ECG basis does not reproduce the cusps of the wave function,
a slow convergence becomes evident for relativistic matrix
elements of the Dirac $\delta$ and the kinetic energy operator $p_a^4$.
To speed up the convergence, the singular operators can be transformed into their equivalent forms,
whose behavior is less sensitive to the local properties of the wave function.
For the Dirac $\delta$ expectation value, such a prescription has been proposed by Drachman \cite{drachman}.
For example, from direct calculation with the basis size of 4096 for $S$-state,
we get $\langle \delta(r_a) \rangle = 35.366\,89 \ldots$, while using the Drachman
regularization approach we improve the convergence by three orders of magnitude (see Table \ref{TBL1}).
Regularization methods have also been applied for the beryllium ground state in the former paper \cite{beqed},
nonetheless the present results are more accurate by 2 orders of magnitude 
due to the better optimized wave function.
For $P$-states, the expectation values of the relativistic and QED operators 
as well as of the Bethe logarithm have been unavailable in literature to date.
Analogous calculations of relativistic terms in the ECG basis have been performed
only for $P$-states of the four-body positronium molecule \cite{Ps2}.
Methods for evaluation of additional integrals ``$1/r^2$'' and ``$1/(r_a r_b)$''
of the form \eqref{eq:fintegral} required for the regularized operators
of the Breit-Pauli Hamiltonian have been developed, resulting in computationally
tractable recursive expressions derived from corresponding master integrals. These
have been presented in the original paper only for three-body systems \cite{gaussintegrals}.
\begin{table}[!htb]
\renewcommand{\arraystretch}{1.3}
\caption{Expectation values of various operators with nonrelativistic wave function for
$2^1S$ and $2^1P$ states of beryllium atom (in a.u.).}
\label{TBL1}
\begin{ruledtabular}
\begin{tabular}{cw{5.13}w{5.13}}
Operator    & \cent{2^1S}  &\cent{2^1P} \\
\hline
${\cal H}_0 $
& -14.667\,356\,498(3) & -14.473\,451\,37(4)\\
$\vec p_a \cdot \vec p_b$
& 0.460\,224\,112(8)   &  0.434\,811\,25(13) \\
$p_a^4$
& 2\,165.630\,1(9)     & 2\,133.321\,1(12) \\
$\delta(r_a)$
& 35.369\,002\,6(6)    &  34.897\,914\,6(8) \\
$\delta(r_{ab})$
& 1.605\,305\,33(9)    & 1.567\,943\,6(2) \\
$p_a^i \bigl(\frac{\delta^{ij}}{r_{ab}}\!+\!\frac{r^i_{ab}\,r^j_{ab}}{r^{3}_{ab}}\bigr) p_b^j$
& 1.783\,648\,19(15)   & 1.624\,185\,8(5) \\
$P(r_{ab}^{-3})$
& -7.326\,766(3) & -7.097\,15(8)\\
$\ln k_0$
& 5.750\,46(2) & 5.752\,32(8) \\
\end{tabular}
\end{ruledtabular}
\end{table}

The calculation of the leading QED corrections is the main challenge of this work.
It is particularly laborious because we deal with the states of the non-vanishing
angular momentum. The explicit form of the $m\,\alpha^5$ terms is given by \cite{araki,sucher}
\begin{eqnarray}
{\cal E}^{(5,0)} &=& \frac{4\,Z}{3}\,\left[\frac{19}{30}+\ln(\alpha^{-2}) - \ln k_0\right]\,
\sum_a\,\langle \delta^3(r_a) \rangle  \\ && \hspace{-1.5cm} +
 \left[\frac{164}{15}+\frac{14}{3}\,\ln\alpha
\right]\,\sum_{a<b} \,\langle \delta^3(r_{ab}) \rangle
-\frac{7}{6\,\pi} \,\sum_{a<b} \,\biggl \langle P\left(\frac{1}{r_{ab}^3} \right) \biggr \rangle  \nonumber.
\label{eq:HQED}
\end{eqnarray}
This expression contains two highly nontrivial terms: the Bethe logarithm $\ln k_0$ 
and the so-called Araki-Sucher distribution $ P (r_{ab}^{-3}) $.
In ECG basis, the latter exhibits exceptionally slow convergence when evaluated directly from its definition.
The regularization is in this case mandatory if one aims at a high accuracy of the final results.
For this purpose we extended the original Drachman's idea and obtained the following
regularized form for the distribution \cite{accelsing}
\begin{eqnarray}
\bigg \langle P\left(\frac{1}{r_{ab}^3}\right) \bigg\rangle &=&
\sum_c \bigg \langle \vec{p}_c\,\frac{\ln r_{ab}}{r_{ab}}\, \vec{p}_c \bigg \rangle  \\ && \hspace{-1cm}  +
\bigg \langle 4 \pi\, (1 + \gamma)\, \delta(r_{ab})
 + 2\, ({\cal E}_0 - V) \,\frac{\ln r_{ab}}{r_{ab}} \bigg \rangle. \nonumber
\label{reg_rabm3}
\end{eqnarray}
As we can see, new classes of the integrals containing factors
of the form ``$\ln r/r$'', ``$\ln r/r^2$'', ``$\ln r_a/(r_a r_b)$'' arise.
With the master integral, such integrals can be expressed
analytically in terms of elementary and Clausen functions.

The evaluation of the Bethe logarithm is the most time consuming part of the calculations.
Formulas for such calculations with the ECG functions have been presented in the former work
devoted to lithium atom \cite{ligauss} and later on applied to the beryllium ground state \cite{beqed}.
In principle, we use the same compact integral representation of Bethe logarithm as before,
however, for the $2^1P$ state such calculations become more sophisticated.
Compared to the ground state, which through the momentum operator 
is coupled only with the virtual $^1P$ states, the $2^1 P$ state requires a complete
set of the $^1S$, $^1P^e$, and $^1D$ intermediate states. These three types of states
can be well represented in the bases $\phi_S$, $\epsilon^{ijk}\,r_a^j\,r_b^k\,\phi_S$, and
$((r_a^i\,r_b^j + r_a^j\,r_b^i)/2 - 1/3\,\delta^{ij}\,r_a^k\,r_b^k)\,\phi_S$, respectively.
Evaluation of $f(t)/\omega$ in the limit of $\omega=0$ is clearly established numerically
from the Thomas-Reiche-Kuhn sum rule for dipole oscillator strengths
$\langle \vec P \,({\cal H}_0-{\cal E}_0)^{-1}\,\vec P \rangle = 3\,Z/2 $.
This value is useful in judging the completeness of the virtual states
and estimation of uncertainties.

Because of principal difficulties, the $m\,\alpha^6$ corrections in their full form were evaluated only for two electron atoms \cite{h6hel}.
Therefore, for the four-electron beryllium atom we use the following approximate formula based on
the hydrogen atom theory \cite{eides}
\begin{eqnarray}
{\cal E}^{(6,0)} &\approx& \pi\,Z^2\,\left[\frac{427}{96} - 2 \ln(2) \right] \sum_a \langle \delta^3(r_a) \rangle.
\end{eqnarray}
This approximation includes the dominating electron-nucleus one-loop radiative correction and neglects
the two-loop radiative, electron-electron radiative, and the higher order relativistic corrections.
On the basis of the experience gained in helium calculations \cite{h6hel}, we estimate,
considering higher charge of the beryllium nucleus, that the neglected
terms contribute less than 20\% to the overall $m\,\alpha^6$ correction.

\begin{table}[!htb]
\renewcommand{\arraystretch}{1.3}
\caption{ Components of the $2^1P - 2^1S$ transition energy and the ionization potential (IP)
for $^9$Be atom in $\icm$. CODATA \cite{codataalpha} inverse fine structure constant
$\alpha^{-1}=137.035\,999\,074(44)$
and the nuclear mass $m_N(^9{\rm Be}) = 9.012\,182\,20(43)$ u \cite{massbe9}. }
\label{TBL2}
\begin{ruledtabular}
\begin{tabular}{cw{8.14}w{6.14}}
Operator    & \cent{2^1P - 2^1S}  & \cent{\mathrm{IP}(2^1S)} \\
\hline
$m\,\alpha^2$          &  42\,557.255(6)   &  75\,190.543(4)\\
$m\,\alpha^2 \,\eta$   & -2.930\,72        &       -4.675\,65    \\
$m\,\alpha^4$          & 12.167(1)         &        7.414\,0(8) \\
$m\,\alpha^5$          & -1.003\,3(14)         &       -0.557\,7(3) \\
$m\,\alpha^6$          & -0.045(9)         &       -0.025(5) \\
Total                  & 42\,565.441(11)   &  75\,192.699(7) \\[2ex]
Theory \cite{chung1993}& 42\,568.80        &                \\
Theory \cite{stankeS}  &                   & 75\,192.667(19)  \\
Theory \cite{beqed}    &                   & 75\,192.510(80)  \\
Experiment \cite{johansson1962}& 42\,565.35(18) &                \\
Experiment \cite{seaton1976} &               & 75\,192.50(10)  \\
Experiment \cite{beigang}    &               & 75\,192.64(6)  \\
\end{tabular}
\end{ruledtabular}
\end{table}

Except for $\mathcal{E}^{(6,0)}$, all the coefficients of the expansion \eqref{eq:expansion}
are evaluated in complete, i.e. no approximation is introduced nor any physical effect
of given order is omitted. Therefore, the uncertainties given in Table~\ref{TBL2} refer to
the incompleteness of the basis set. 
On the basis of our former work \cite{lit_rel} on Be$^+$, 
the higher order in $\alpha$ and $\eta$ contributions, 
namely $m\,\alpha^2 \eta^2$ and $m\,\alpha^4 \eta$ are estimated to be less
than $0.001~\icm$ to both the transition energy 
and ionization potential, and thus they are negligible when compared to the present
uncertainty of $0.01~\icm$. We note in passing that
in Table~\ref{TBL2} for the values without the uncertainty all the quoted digits are certain.
In evaluation of the IP value we used the ground state energy level of the beryllium cation
$E(\text{Be}^+)=-14.325\,836\,7$ a.u. calculated with Hylleraas wave functions \cite{lit_rel}.

The accuracy of $0.011~\icm$ for the transition $2^1P - 2^1S$ and $0.007~\icm$
for the ionization energy has been achieved due to the recent progress made in two directions.
The first one, essential to reach this accuracy, is the improvement
in the optimization process of the nonrelativistic wave functions leading to
the overall increase in numerical precision. The second direction is the complete
treatment of the leading relativistic and QED effects. More specifically,
the approach to effectively calculate the many electron Bethe logarithm and mean values
of singular operators, like the Araki-Sucher term, has been developed \cite{beqed}.
Particularly, an extension of the numerical methods for relativistic and QED corrections
on $P$-states of a four-electron system is presented here.

The $m\,\alpha^5$ and $m\,\alpha^6$ terms involve
the interaction of the electrons with the vacuum fluctuations
of electromagnetic field, the electron-positron virtual pair creation,
and the retardation of the electron-electron interaction.
Results of Table~\ref{TBL2} show clearly that taking into account 
such energetically subtle QED effects
is unavoidable in to reach the agreement between the experiment and the theory
and that it enables testing of QED.
For example, the overall contribution of the QED effects to the $2^1P - 2^1S$ transition
energy amounts to $1.048(9)\ \icm$ and is an order of magnitude higher
than the experimental uncertainty.
Currently, the accuracy reached by theory for the transition frequencies exceeds
by an order of magnitude that of the known  measurements for beryllium atom.
At this level of accuracy we are able to resolve
the $2^1P - 2^1S$ line discrepancy of $3.45~\icm$
between the experiment \cite{chung1993} and theory in favor of the latter.
Although, the available semi-empirical results of the ionization energy \cite{beigang}
agree well with the more accurate theoretical results obtained here, we hope that
the increased level of accuracy of the theoretical predictions
will be a stimulus for new, more accurate measurements.

Uncertainty of our results comes mainly from the neglect of the higher order relativistic and QED
corrections of the order $\alpha^{6,7}$. Evaluation of these term sets the direction of our
future efforts.
Also the numerical accuracy of the nonrelativistic energy has to be improved to achieve
further progress in theoretical predictions.

The recursive method of evaluation of the integrals \eqref{eq:fintegral}
employed in this work allows an application of the ECG functions
to the states with non-vanishing angular momentum. It establishes a framework for a
high accuracy studies of the fine structure and the hyperfine splitting in the beryllium atom.
The isotope mass shifts can also be precisely calculated. However, an accurate experimental
data are necessary to enable an extraction of a nuclear-model-independent
charge radii from isotope shifts by combining high-accuracy measurements with atomic theory.
This is of special interest for the halo nuclei (e.g. $^{11}$Be) for which
analogous results obtained recently from the $2^2P - 2^2S$ transition
in beryllium cation \cite{yan_rel,lit_rel} require a confirmation.
Systematic extension to transition energies involving $D$-states is mandatory
to resolve other severe discrepancies between theory and measurements
pointed out by Chung and Zhu \cite{chung1993} like e.g. the largest one
of $7.38~\icm$ for the $3^1D-2^1S$ transition.
To our knowledge, no precise calculations of the Bethe logarithm with
such nontrivial angular momentum structures have been performed for any few-electron
systems so far (even for helium). The methodology presented in this paper
opens up a route towards removing such obstacles and, what is more,
is very promising in applications to five- and six-electron systems.

\section*{Acknowledgments}
This research was supported by the NCN grant 2011/01/B/ST4/00733 and by a computing grant
from Pozna\'n\ Supercomputing and Networking Center, and by PL-Grid Infrastructure.


\begin{thebibliography}{99}

\bibitem{johansson1962} L. Johansson, Ark. Fys. {\bf 23}, 119 (1962).
\bibitem{bozman1953} W. R. Bozman, C. H. Corliss, W. F. Meggers, and R. E. Trees, J. Res. Natl. Bur. Stand. (U.S.) {\bf 50}, 131 (1953).
\bibitem{seaton1976} M. J. Seaton, J. Phys. B {\bf 9}, 3001 (1976).
\bibitem{beigang} R. Beigang, D. Schmidt, and P. J. West, J. Phys. (Paris), Colloq. {\bf 44}, C7-229 (1983).
\bibitem{beqed} K. Pachucki and J. Komasa, Phys. Rev. Lett. {\bf 92}, 213001 (2004).
\bibitem{stankeS} M. Stanke, D. K\c{e}dziera, S. Bubin, and L. Adamowicz, Phys. Rev. A {\bf 75}, 052510 (2007Í?).
\bibitem{bushaw2007} B. A. Bushaw, W. N{\" o}rtersha{\"u}ser, G.W. F. Drake, and H.-J. Kluge, Phys. Rev. A {\bf 75}, 052503 (2007).
\bibitem{sansonetti2011} C. J. Sansonetti, C. E. Simien, J. D. Gillaspy, J. N. Tan, S. M. Brewer, R. C. Brown, S. Wu, and J. V. Porto, Phys. Rev. Lett. {\bf 107}, 023001 (2011)
\bibitem{yan_rel} Z.-C. Yan, W. N\"ortersh\"auser and G.W.F. Drake, Phys. Rev. Lett. {\bf 100}, 243002 (2008); ibid. {\bf 102}, 249903(E) (2009).
\bibitem{lit_rel} M. Puchalski and K. Pachucki, Phys. Rev. A {\bf 78}, 052511 (2008).
\bibitem{komasa2011} J.~Komasa, K.~Piszczatowski, G.~Lach, M.~Przybytek, B.~Jeziorski, and K.~Pachucki, J. Chem. Theory Comput. {\bf 7}, 3105 (2011).
\bibitem{przybytekHe2} M. Przybytek, W. Cencek, J. Komasa, G. Lach, B. Jeziorski, and K. Szalewicz, Phys. Rev. Lett. {\bf 104}, 183003 (2010).
\bibitem{h6hel} K. Pachucki, Phys. Rev. A {\bf 74}, 022512 (2006); ibid. {\bf 74}, 062510, (2006); ibid. {\bf 76} 059906(E) (2007).
\bibitem{PY10} K.~Pachucki and V.~A.~Yerokhin, Phys. Rev. Lett. {\bf 104}, 070403 (2010).
\bibitem{chung1993} K. T. Chung, X.-W. Zhu, and Z.-W.Wang, Phys. Rev.A {\bf 47}, 1740 (1993).
\bibitem{drakeblH} G. W. F. Drake, Can. J. Phys. {\bf 66}, 586 (1988).
\bibitem{drakelnk0} G. W. F. Drake and R. A. Swainson, Phys. Rev. A {\bf 41}, 1243 (1990).
\bibitem{bubinP} S. Bubin and L. Adamowicz, Phys. Rev. A {\bf 79}, 022501 (2009).
\bibitem{sharkeyD} K. L. Sharkey, S Bubin and L. Adamowicz, Phys. Rev. A {\bf 84}, 044503 (2011).
\bibitem{harris} F. E. Harris, Int. J. Quant. Chem., {\bf 106}, 3186â??3189 (2006).
\bibitem{gaussintegrals} K. Pachucki and J. Komasa, Chem. Phys. Lett. {\bf 389}, 209 (2004).
\bibitem{bethe} H. A. Bethe and E. E. Salpeter, {\it Quantum mechanics of One- and Two-electron Atoms} (Springer, Berlin, 1957).
\bibitem{drachman} R. J. Drachman, J. Phys. B {\bf 14}, 2733 (1981).
\bibitem{Ps2} M. Puchalski and A. Czarnecki, Phys. Rev. Lett. {\bf 101}, 183001 (2008).
\bibitem{araki} H. Araki, Prog. Theor. Phys. {\bf 17}, 619 (1957).
\bibitem{sucher} J. Sucher, Phys. Rev. {\bf 109}, 1010 (1958).
\bibitem{accelsing} K. Pachucki, W. Cencek and J. Komasa, Phys. J. Chem. Phys. {\bf 122}, 184101 (2005).
\bibitem{ligauss} K. Pachucki and J. Komasa, Phys. Rev. A {\bf 68}, 042507 (2003).
\bibitem{eides} M. I . Eides, H. Grotch, and V. A. Shelyuto, Phys. Rep. {\bf 342}, 63 (2001).
\bibitem{massbe9} G. Audi, A.H. Wapstra, and C. Thibault, Nucl. Phys {\bf A729}, 337 (2003); URL http://amdc.in2p3.fr /web/masseval.html.
\bibitem{codataalpha} P. J. Mohr, B. N. Taylor, and D. B. Newell, Rev. Mod. Phys. {\bf 84}, 1527 (2012).


\end{thebibliography}
\end{document}